\documentstyle[12pt,aaspp4]{article}
\input epsf
\epsfverbosetrue
\def\msat{m_{\rm sat}}
\def\peri{R_{\rm peri}}
\def\apo{R_{\rm apo}}
\def\trad{T_{\rm R}}
\def\tsim{T_{\rm sim}}

\begin{document}

\title{Interpreting Debris from Satellite Disruption In External Galaxies}

\author{
Kathryn V. Johnston\altaffilmark{1},
Penny D. Sackett\altaffilmark{2} and
James S. Bullock\altaffilmark{3}}

\altaffiltext{1}{Van Vleck Observatory, Wesleyan University,
Middletown, CT 06459; kvj@astro.wesleyan.edu}
\altaffiltext{2}{Kapteyn Astronomical Institute, University of
Groningen, 9700 AV Groningen, Netherlands; psackett@astro.rug.nl}
\altaffiltext{3}{Department of Astronomy, The Ohio State University, 
140 W 18th Ave, Columbus, OH 43210-1173; james@astronomy.ohio-state.edu}

\begin{abstract}
We examine the detectability and interpretation of debris trails caused by
satellite disruption in external galaxies using semi-analytic approximations
for the dependence of streamer length, width and surface brightness on
satellite and primary galaxy characteristics. The semi-analytic method is
tested successfully against N-body simulations and then applied to three
representative astronomical applications. First, we show how streamer
properties can be used to estimate mass-to-light ratios $\Upsilon$ and streamer
ages of totally disrupted satellites, and apply the method to the stellar arc
in NGC~5907.  Second, we discuss how the lack of observed tidal debris around
a satellite can provide an upper limit on its mass-loss rate, and, as an
example, derive the implied limits on
mass-loss rates for M32 and NGC~205 around
Andromeda.  Finally, we point out that a statistical analysis of streamer
properties might be applied to test and refine
cosmological models of hierarchical galaxy formation, and use the predicted
debris from a standard $\Lambda$CDM realization 
to test the feasibility of such a study. 
Using the Local Group satellites and the few known examples of debris trails
in the Galaxy and in external systems, we estimate that the best current
techniques could 
characterize the brightest ($R < 29$~mag/ arcsec$^{2}$) portions of the
youngest (3 dynamical periods) debris streamers. If systematics can be
controlled, planned large-aperture telescopes such as CELT and OWL may allow
fainter trails to be detected routinely and thus
used for statistical studies such as those required for tests of galaxy
formation.
\end{abstract}


\section{Introduction}

In hierarchical models of galaxy formation, galaxies are built up from
many smaller pieces. 
Theoretical work has shown that the
destruction of a dwarf galaxy that is much smaller than its 
parent system is likely to leave observable signatures in phase space, with
debris from the accreted dwarf simply phase-mixing along its original
orbit (\cite{t93}; \cite{j98}; \cite{hw99}).
The mixing time depends on the dwarf's mass and orbit, and is shortest for the
most massive dwarfs in the inner galaxy.
In this case the debris can become difficult to detect spatially within a few
Gyrs of its creation, but remains dynamically cold and identifiable
in the form of moving groups of stars (\cite{hw99}). Several examples of
such groups have been identified in our own Milky Way (\cite{mmh96}; \cite{h+99}).
In the outer galaxy the debris forms streams that can maintain coherence for
many Gyrs (\cite{jhb96}) and whose evolution can be described approximately
using simple analytic estimates (\cite{t93}; \cite{j98}).
The prime example of such debris that has been found from its overdensity in
star counts rather than kinematics is that associated with the Sagittarius
dwarf galaxy (\cite{mom98}; \cite{m+99}; \cite{i+00}).
In addition, recent results from SSDS on the magnitude distribution of
A-colored stars (\cite{sdss00a}) and RR Lyrae variables (\cite{sdss00b})
suggest that the Milky Way's stellar halo becomes quite lumpy beyond
20--30~kpc.  Bullock, Kravtsov \& Weinberg (2001)
have used semi-analytic galaxy formation
models to show that such substructure is consistent with the hierarchical view
of structure formation.

The multitude of recent theoretical and observational results on streamers
in the Milky Way naturally leads to the question of what we might
expect to see around external galaxies. 
The latter have the advantage of being non-unique 
(and thus offer the opportunity of building a sample of
results to evaluate ``typical'' galaxy properties), 
but the disadvantage of being restricted to streamers seen from a single random
viewing angle.  In external galaxies, only surface brightnesses 
and line-of-sight
velocities are available; Galactic streamers offer
the potential for full phase-space information.

One might ask, for example, whether the debris from the Sagittarius dwarf (Sgr)
would be observable from a viewpoint outside the Milky Way.
Ibata et al.\ (2000) have reported finding 38
faint, high-latitude, Carbon stars in their survey covering 
Galactic latitudes $|b|>30^\circ$ that lie within 10~degrees
of the projected orbit of Sgr.
By appropriately 
scaling the \cite{whitelock96} survey for Carbon stars in the central 
regions of Sgr they calculate $\sim 100$ Carbon stars should lie
in the dwarf itself.
Since Sgr is estimated to have a total $V$ luminosity of $\sim$$2\times
10^7 L_{V, \odot}$ (\cite{m98}) and presently
be on an orbit oscillating between $\sim$$15 \,$kpc and $\sim$$60 \,$kpc,
the Carbon star result suggests a streamer of debris ejected in the last
few Gyrs (Carbon stars have intermediate ages) with a
total luminosity of $\sim$$2\times10^7 L_{V, \odot}$ and physical dimensions
of $\sim$5~kpc by $\sim$250~kpc.
Viewed from a distant external galaxy, such a streamer would have an
average surface brightness of $\sim$$\mu_V = 31~$V~mag/arcsec$^2$.
This rough calculation is in agreement with
estimates of $\mu_V$=30-31 for the surface brightness of debris over 
30 degrees away from the main body of Sgr 
mag/arcsec$^{2}$, from both main sequence (\cite{mom98}) and giant star 
(\cite{m+99}) overdensities in the region.
In addition, a 
recent paper (\cite{delgado01}) reports a similar surface brightness for a 
population of star possibly associated with Sgr in the Northern Galactic 
Hemisphere over 50kpc from the Galactic center.
Since streamer luminosity is not smoothly distributed,
but instead has the largest surface brightness at apocenter,
the brightest portions of debris trails like that of Sgr
around an external galaxy are likely to be just at current limits of detection 
(see \S\S 3, 5).
Many examples of low surface brightness features already have been
detected around other galaxies (\cite{mh97}).
A particularly fine example is the streamer in NGC~5907 (\cite{s+98}) which
we discuss further in \S 4.2.1.

The combination of
observations of substructure in the Milky Way's stellar
halo with the existence of
features around external galaxies motivates us to
apply theoretical work on the evolution of
debris about the Milky Way and ask what we might expect to see 
around other galaxies.
We review the relevant dynamics and present analytic expressions
for the evolution of the geometry of tidal streams in \S 2.
In \S 3 we discuss under what circumstances such streams should
be observable by comparing our prediction for their surface
brightness with the background contamination from their parent
galaxy's disk, bulge and halo light.
In \S 4 we apply our results to: interpreting the debris seen around
NGC 5907; placing limits on the mass loss rate from satellites of
M31; and exploring whether the number of features seen around
external galaxies might be of interest in the context of
hierarchical models of galaxy formation.
We discuss future observational prospects in \S 5 and conclude in \S 6.

\section{Debris Dynamics and Theoretical Interpretation of Young Streamers}

In this section we develop analytic expressions
for the morphology of tidal streamers and compare them
with the results of simulations of satellite galaxy destruction.
In all these simulations, with the exception of
those used for the blue and green points in Figure \ref{wvst.fig},
the parent galaxy potential was fixed, with
bulge, disk and halo components as described in the work of
Johnston, Hernquist \& Bolte (1996).
In the other cases the halo potential was replaced by a flattened
analog that generated the same rotation curve in the disk
plane but had axis ratio $q=0.7$ in density.
In all cases, the satellite galaxy was represented by
particles initially distributed as a Plummer model,
whose mutual interactions were calculated using
a basis function expansion code (\cite{ho92}).
Table \ref{simstab} summarizes the satellite and orbital properties
used in the simulations and the papers in which they originally appeared.

\begin{table}
\begin{center}
\begin{tabular}{cccccccc}
Figure	& $\msat$	 & $s$ & $\peri$ & $\apo$ & $\trad$ & $\tsim$ & 
Paper \\
	& $M_\odot$	 &     & kpc	  & kpc    & Gyr    & Gyr    \\
\hline
\hline
\ref{xyej.fig}, \ref{wvsr.fig}	& $10^8$ & 0.08 & 30     & 450    & 6.0     &	30.0  
& JCG01 (Model 4) \\
\hline
\ref{wvst.fig}, yellow	& $10^6$ & 0.015 & 26 & 37 & 0.7 &  7.0 & this paper \\
\hline
\ref{wvst.fig}, red	& $10^9$ & 0.15 & 26 & 37 &  0.7 &  7.0 & this paper  \\
\hline
\ref{wvst.fig}, blue & $10^7$ & 0.029 & 45 & 135& 2.2 & 19.7 & this paper \\
\hline
\ref{wvst.fig}, green & $10^7$ & 0.034 & 25 & 60& 1.2 & 19.7 & this paper \\
\hline
\ref{surf.fig}a	& $10^9$ & 0.10	& 40	& 135 & 2.0 & 10 
& JHB96 (Model 1) \\
\hline
\ref{surf.fig}b	& $4\times 10^7$ & 0.04 & 41 & 175 & 2.7 & 10
& JHB96 (Model 2) \\
\hline
\ref{surf.fig}c 	& $2\times 10^7$  & 0.03 & 45 & 144 & 1.4 & 10 
& JHB96 (Model 3) \\
\hline
\ref{surf.fig}d 	& $7\times 10^6$ & 0.07	& 20 & 140 & 0.9 & 10 
& JHB96 (Model 4) \\
\end{tabular}
\end{center}
\caption{
\label{simstab}
Parameters of satellites and orbits for the simulations used to
test and illustrate our results in Figures 1--5 (with panels
in Fig.~5 indicated by letters a--d from top to bottom).
References in the last column: JCG01 -- \cite{jcg01}; JHB96 -- \cite{jhb96}.}
\end{table}

Tremaine (1993) and Johnston (1998)
outline a simple physical picture for how debris
from the destruction of a satellite orbiting in a near-spherical
potential disperses along the satellite's path.
Here we extend their approach to describe the full
geometry of tidal streamers.
We expect our results to be valid when
tidal streamers are dynamically young and still distinct in 
coordinate space ---
for example when following the dispersal of
debris from small satellites that are disrupting
in the near-spherical, outer halo of a galaxy
where the timescale to mix fully along an orbit is
longer than the age of the Universe. (We discuss what
we mean by ``small'' and ``outer halo'' more fully in
\S 2.3.3.) 
Helmi \& White (1999) give a more rigorous treatment of
debris dispersal applicable to a satellite orbiting in a static potential
of any geometry and capable of following the full phase-space distribution of
tidal debris as it becomes well-mixed --- for example, to
describe signatures of ancient accretion events in the
inner halo.

In all equations below, the first equality
holds for a tidal streamer in a general spherical potential
$\Phi(R)$ and the second expression (following the arrow)
is for the specific case
of a logarithmic potential $\Phi=v_{\rm circ}^2 {\rm log}(R)$, 
as might be appropriate for a galaxy with a flat rotation curve
and circular velocity $v_{\rm circ}$.
The equations can be applied in the regime in which
the {\it tidal scale} $s \ll 1$, where
\begin{equation}
	s\equiv \left({m \over M(R_p)}\right)^{1/3}
	\, \longrightarrow \, 
	\left({Gm \over v_{\rm circ}^2 R_p}\right)^{1/3},
\label{tscale}
\end{equation}
$m$ is the mass of the satellite, $M(R)=-(R^2/G)d\Phi/dR \longrightarrow
v_{\rm circ}^2 R/G$ 
is the mass of the parent
enclosed within radius $R$, and $R_p$ is the pericenter of
the satellite's orbit
(where mass loss predominantly occurs).

\subsection{Dispersal Along the Satellite's Path}
 
The physical scale at which a satellite loses mass is given by
its {\it tidal radius\/} 
\begin{equation}
\label{rtide}
	r_{\rm tide} =	s\, R_p.
\end{equation}
Thus, the specific orbital energies and angular momenta
of the stars in the debris will be distributed 
about the satellite's own orbital energy $E$ and angular momentum
$J$ with a characteristic internal energy scale
\begin{equation}
\label{epsilon}
        \epsilon = r_{\rm tide}  \left[{d\Phi \over dR}\right]_{R_p}
	\, \longrightarrow \, 
	s v_{\rm circ}^2
\end{equation}
and angular momentum scale
\begin{equation}
\label{jay}
	j=sJ~.
\end{equation}

The time periods of orbits in a spherical potential depend
more strongly on energy than on angular momentum
(see e.g., Johnston 1998),
so it is primarily the spread in energy $\epsilon$ within the debris that is
responsible for the length of
streamers seen in simulations of
tidal disruption.
After a time $t$, the debris will spread an angular distance of order
\begin{equation}
\label{psi}
	\Psi\sim 4 \epsilon 
\left[{1 \over T_{\Psi}}
{\partial{T_\Psi} \over \partial{E}}\right]_{J=J_{\rm circ}} {2\pi t
\over T_{\Psi}}
	\, \longrightarrow \,4 s {2\pi t \over T_{\Psi}}
\end{equation}
along the satellite's orbit, where $T_{\Psi}$ is the
azimuthal time period of a circular orbit of energy $E$.
The factor of four reflects 
the fact that debris energies are predominantly spread
in the range $-2\epsilon < \Delta E < 2 \epsilon$
around the satellite's orbital energy (see Johnston 1998 and \S2.2 below).

The discussion above indicates that once a set of stars
becomes unbound from the satellite the constituents sort themselves
in energy along a streamer.
Hence, debris at 
a particular point along the streamer will have an average
energy $\Delta E$ relative to the satellite's orbital
energy and will be
offset correspondingly in radius from the satellite's path by
\begin{equation}
\label{dr}
	\Delta R \sim \Delta E {d R \over d\Phi} 
	\, \longrightarrow \, {\Delta E \over v_{\rm circ}^2} R~.
\end{equation}

The concepts encapsulated in equations (\ref{epsilon}) and (\ref{psi}) 
are illustrated in the left-hand panels of
Figure \ref{xyej.fig}.
These show the positions
in the orbital plane of particles at three points during
a satellite disruption simulation.
In order to emphasize the evolution of a debris population with age, 
only those particles lost on the first pericentric passage
of the satellite are plotted. The panels correspond to
the first, third and fifth apocenter following this
(hence when the debris is 
``age'' 0.5, 2.5 and 4.5 $T_R$).
The particles are
color coded by orbital energies relative to the satellite's orbit.
The color coding clearly shows the particles sorting themselves in 
energy along the satellite's orbit, with the extent of the debris
along the orbit steadily increasing with time.

\subsection{Dispersal Perpendicular to the Satellite's Path}

In contrast to the orbital time periods,
the precession rate ($T_R/T_\Psi$) of the turning points of an
orbit in a spherical potential depends primarily
on the angular momentum of the orbit.
Hence the spread in the angular momentum in the
debris will cause the angular positions of the turning points of
stars in the debris to spread out over time.
This will lead to an increase in the streamer's 
width in the orbital plane of order
\begin{eqnarray}
\label{wt}
	w(R)&=&w_0+ 2jR \left({\partial{T_R/T_{\Psi}}
\over \partial{J}}\right) {2\pi t\over T_{\Psi}} =
	w_0 +  R\left[J {\partial{T_R/T_{\Psi}} 
\over \partial{J}}\right] s {4\pi t\over T_{\Psi}} \cr
	&=& w_0 \left(1+\left[J {\partial{T_R/T_{\Psi}} 
\over \partial{J}}\right] {4\pi t\over T_{\Psi}}\right) \cr
	&\, \longrightarrow \,& w_0+R\left[J {\partial{T_R/T_{\Psi}} 
\over \partial{J}}\right]  {\Psi \over 2},
\end{eqnarray}
where
\begin{equation}
\label{width}
	w_0=w_0(R)=s R
\end{equation}
is the width of the streamer at $R$ set by the initial spread in the
turning points due to the tidal radius at pericenter.
The factor of two prefacing the second term after the first equality 
reflects the fact
that although the full range in angular momenta in the debris 
is of order $-2j < \Delta J < 2j$ (analogous to the energy range
described in the previous section), 
the spread at any point along the trail is of order $2 j$
since negative (positive) angular momentum differences
correspond to debris in leading (trailing) streamers.

\begin{figure}
\begin{center}
\plotone{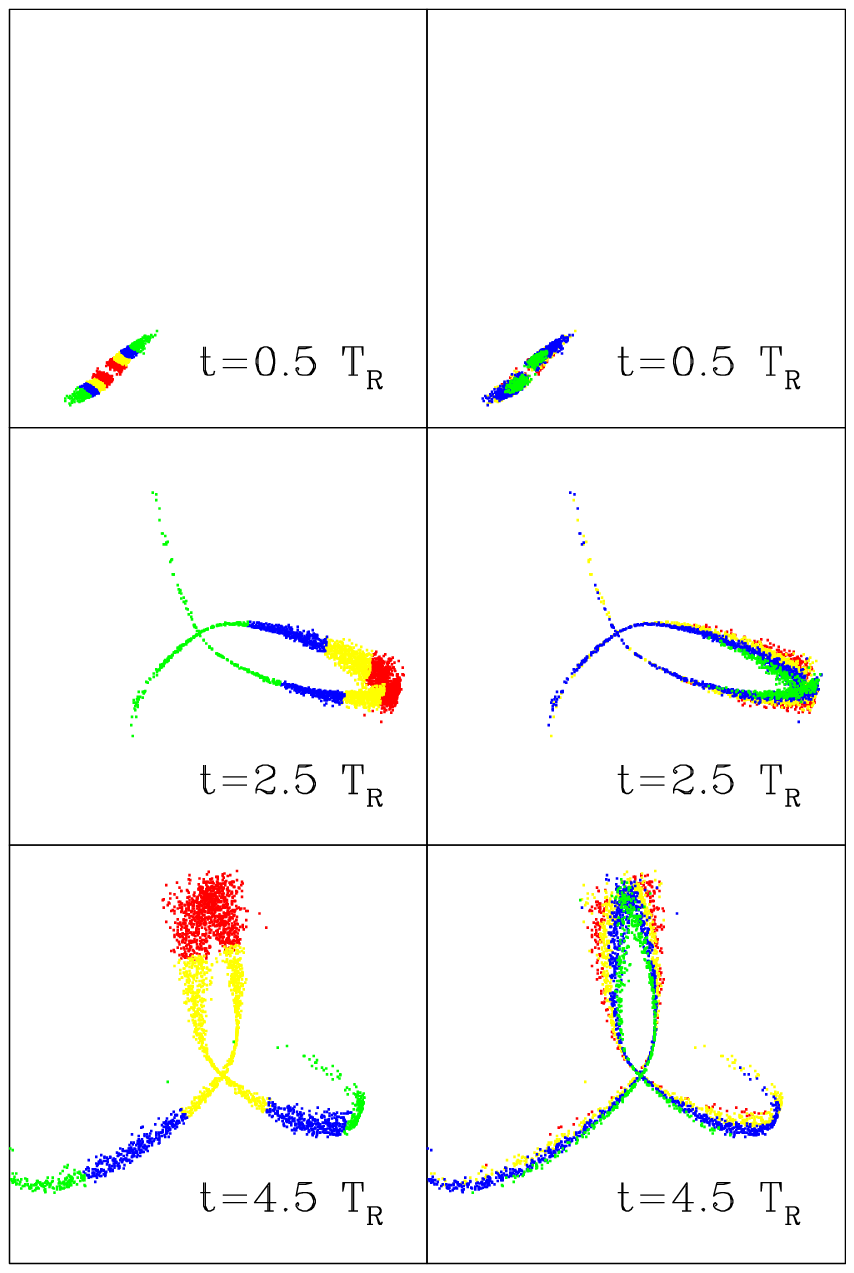}
\caption{Positions
in the orbital plane of particles at the end of a simulation
of satellite disruption, color coded first by their
orbital energies (left-hand panels) and then orbital angular
momenta (right-hand panels) relative to the satellite's orbit
(the box size is 1000~kpc on a side).
The red, yellow, blue, and green colors represent changes (scaled by
$\epsilon$ and $j$) in
the range $0.5-1.25, 1.25-1.75, 1.75-2.25$, and $2.25-3$, respectively.
\label{xyej.fig}}
\end{center}
\end{figure}

Note that this description naturally explains the clear increase in the
width of the streamers with $R$ in each of the panels of
Figure \ref{xyej.fig}: an angular offset between the turning points of
debris particle orbits corresponds to a physical width that increases
with $R$. 
The right-hand panels of this figure repeat the plots in the left-hand panels,
but are now color coded according to the angular momenta of the particles
relative to the satellite's orbit.
This coding illustrates how the particles sort themselves parallel to
the orbit in angular momenta.

\begin{figure}
\begin{center}
\plotone{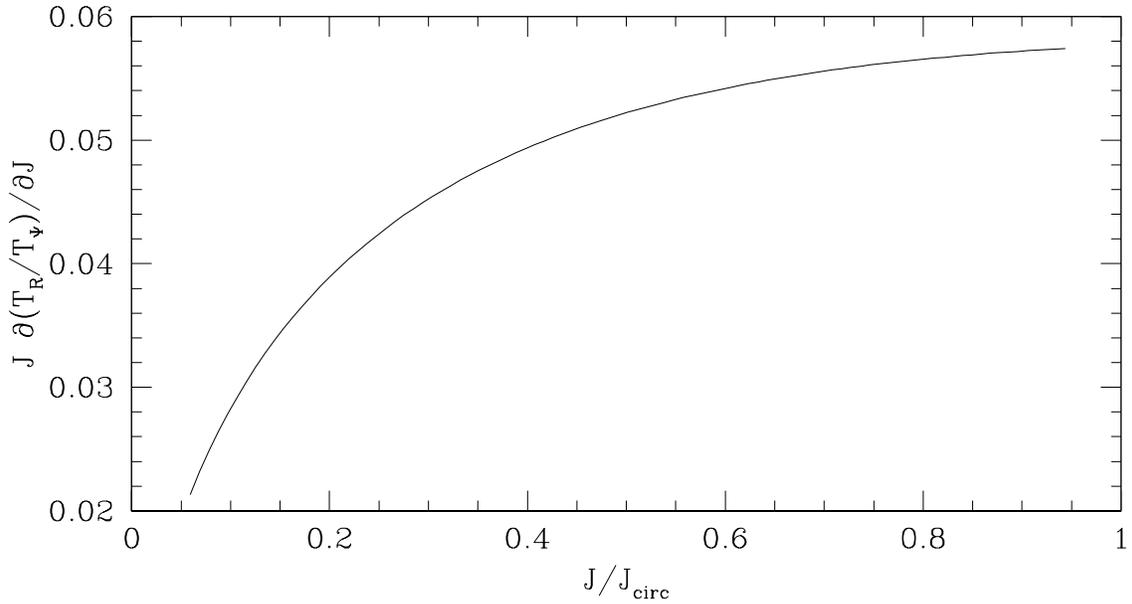}
\caption{The dependence of $J \, \partial (T_R/T_{\Psi})/ \partial J$
(the quantity in square brackets in eq.\ (\ref{wt}))
on angular momentum for orbits in a scale-free logarithmic
potential. In this case the quantity is independent of energy.
\label{dtdj.fig}}
\end{center}
\end{figure}

Since it is hard to visually assess the increase of the width of the 
streamers with time in Figure~\ref{xyej.fig},
the quantity in square brackets in equation (\ref{wt}) is plotted 
in Figure~\ref{dtdj.fig} as a function of angular momentum for
a logarithmic potential, demonstrating that it is
typically of order 0.05.
This implies that the age $t$ of the streamer must be
$t\sim (1/0.05)(T_\Psi/4 \pi) \sim
1.5 T_\Psi \sim 2.25 T_R$ or larger (where the last equality comes from
the range of precession rates in a logarithmic potential 
$0.6 < T_R/T_\Psi <0.7$) before the streamer's
width starts increasing significantly beyond $w_0$,
and that the streamer length $R \Psi$ will always be
at least 10 times its width $w$.

Of course, in most instances the line of sight will not be perpendicular
to the plane of a satellite's orbit, so we must assess
how debris will spread perpendicular to the satellite's motion.
In a spherical potential, the planar nature of the orbits guarantees
that the debris will not spread perpendicular to the plane of the
orbit beyond the height set at pericenter
\begin{equation}
\label{height}
        h(R) \sim \left({r_{\rm tide} \over R_p}\right) R = s\,R
\equiv w_0(R).
\end{equation}
In non-spherical potentials, $h$ will
increase with time due to the precession of the orbital
plane in a manner analogous to the increase in width due
to the precession of turning points in the orbital plane (\cite{hw99}).

We test the expectations outlined in this section more thoroughly in \S 2.3.1.

\subsection{Tests of the Analytic Description}

\subsubsection{Evolution of Width and Height of Streamers}

In this section we present plots of the width $w$ and height $h$ of streamers 
from a variety of simulations of satellite disruption,
measured at several points along their length and at many different times.
Each measurement was made at the position of a randomly-selected
debris particle; the instantaneous
plane of the orbit was defined at that point by the test particle's
position and motion.
All particles lost during the same pericentric passage 
and located within $r_{\rm max}=0.25 R/R_{\rm peri}$ of
the test particle were labeled and ordered in distance
from the orbital plane.
The height $h$ was defined as the difference in the distances
of the particles at the 10th and 90th percentile of the
ordered set.
To find a corresponding width, the above process was repeated for
distances along axes defined in the orbital plane and centered on
the test particle.
Since the exact orientation of the stream was unknown,
the results from 30 equally-spaced axis orientations were compared, and
$w$ was taken to be the minimum of all these measurements.

\begin{figure}
\begin{center}
\plotone{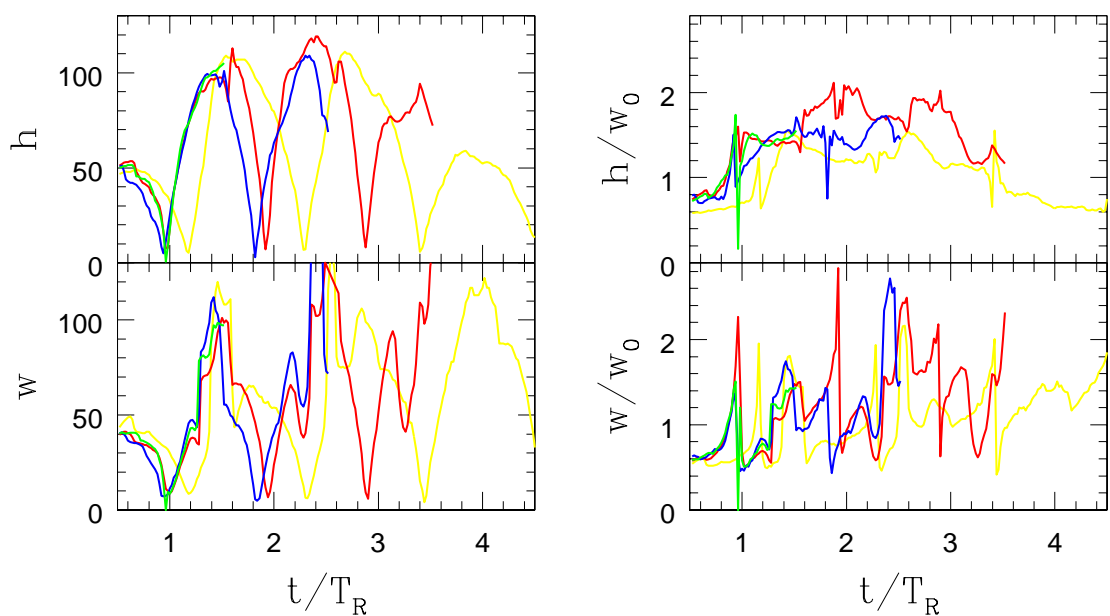}
\caption{The left-hand panels present width and height measurements
of debris as a function of time.
The measurements were made around four 
particles lost on consecutive pericentric 
passages in a simulation.
The right-hand panels scale these measurements by the expected initial
height and width, $w_0$, given by equation (\ref{width}).
\label{wvsr.fig}}
\end{center}
\end{figure}

The left-hand panels of Figure \ref{wvsr.fig} illustrate the results of this
analysis by plotting the evolution of $h$ and $w$ with time as measured around
four test particles, each lost during different pericentric passages of the
same simulation used in Figure \ref{xyej.fig}.
Time since the test particle was lost is expressed in units of
the satellite's radial orbital time period. 
The order-of-magnitude, periodic swings in the sizes of $h$ and $w$ 
are a reflection of the highly eccentric orbit of
the satellite (which the debris closely tracks), with
the smallest/largest values corresponding to the 
pericenters/apocenters of the particle's orbit.
Note that the increasing discrepancy between the time of these 
turning points is another signature of the debris drifting apart into
streamers once it is lost.

The right-hand panels of Figure \ref{wvsr.fig} present the
same information as the left-hand panels,
but are normalized at each point by the expected initial width,
$w_0$, given in equation (\ref{width}).
This demonstrates that much of the behavior of
the width and height of 
the debris is captured by this simple expression over the first few 
dynamical periods following disruption.

We tested this statement further using four more simulations with different
satellite masses, orbits and parent galaxy potentials.
Figure~\ref{wvst.fig} displays the average of $h/w_0$ and
$w/w_0$ measured around 10 test particles
in debris populations lost on the first pericentric passage of each simulation
(with the error bars indicating the dispersion) as a function of the
time since that passage.
The red and yellow points are for satellites on the same orbits, orbiting
the same galaxy (with a spherical halo component), but
with masses $10^9 M_\odot$ and $10^6 M_\odot$ respectively.
These satellites have tidal scales $s$ that differed by a factor of 10,
yet the points in the plot lie almost on top of each other.
Note also that both $h$ and $w$ are increasing gradually with time
because, even though the halo was perfectly spherical, 
the chosen orbit was fairly small (see Table 1) and
the disk component of the parent galaxy caused some
precession of the orbital plane.
The blue and green points in the figure are for
satellites of the same mass on two different
orbits in a galaxy with an oblate halo.
Here, the precession of the orbital plane is much more pronounced,
which is reflected in faster growth of the width and height.
We conclude that as long as the debris is no older than
$t = 4 T_R$, equations (7) and (9) adequately describe
the evolution of streamer height and width for a wide range
of tidal scales, halo oblateness and orbital parameters.
 
\subsubsection{Length and Surface Brightness Predictions}

Johnston (1998) used the ideas encapsulated in equations
(\ref{epsilon}) and (\ref{psi})
to understand the 
characteristics of debris dispersal
seen in N-body simulations of the disruption of spherical stellar
systems along a variety of orbits in a potential chosen to
represent the Milky Way.
She found that the debris in each simulation
followed the same distribution in energies relative to the
satellite's orbit when scaled by the factor given in equation (\ref{epsilon}) 
over the range $-3\epsilon < \Delta E < 3\epsilon$
(and predominantly in range $-2\epsilon < \Delta E < 2\epsilon$ as noted
above).
Using this intrinsic distribution, she developed a
semi-analytic method that could
successfully reproduce the mass density 
along the tidal tails seen in any simulation, given the mass,
orbit and mass loss rate of the satellite.
This success confirms the predictive power of
expressions in equations (\ref{epsilon}) and (\ref{psi}).

We tested the additional scalings derived above
against the simulations of Johnston, Hernquist \& Bolte (1996)
by using the same methods
to find the mass density and energy distribution
along the streamers, and equations (\ref{width}), (\ref{height}) 
and (\ref{dr}) to predict the width and height of the streamers and their
offset from the satellite's orbit. 
The right-hand panels of
Figure~\ref{surf.fig} show the surface density map produced using this
semi-analytic method. The left-hand panels show the 
corresponding map produced from the actual final positions of particles
in the simulations. The model successfully reproduces the 
position, geometry and surface density of the streamers produced by  
the simulations.
In particular, note that at the point along the streamers where
the surface brightness drops by roughly a magnitude from its maximum, the
debris stars have $\Delta E \sim \pm 2 \epsilon$, confirming 
the factor of four in equation (\ref{psi}) for our
estimate of the streamer length.

\subsubsection{Approximations Adopted}

\begin{figure}
\begin{center}
\plotone{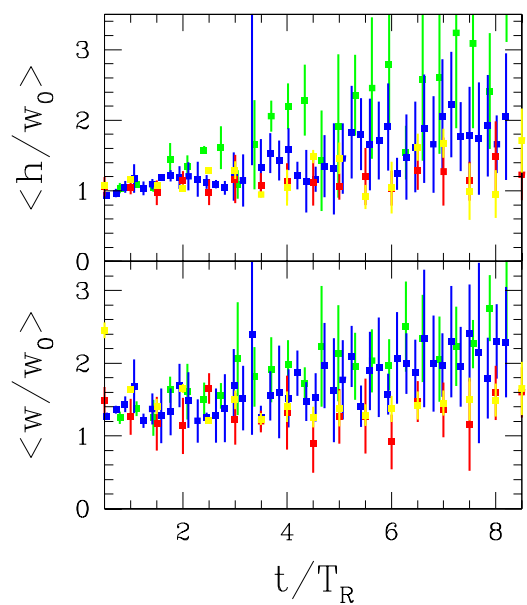}
\caption{Average (points) and dispersion (error bars)
over scaled height and width measurements made around 
ten randomly chosen particles as a function of time.
The yellow and red points are for satellites of masses
$10^6M_\odot$ and $10^9 M_\odot$ respectively on the same orbit in
a galaxy with  spherical halo.
The blue and green points are for a satellite of mass $10^7 M_\odot$
on two different orbits in a galaxy with an oblate halo.
\label{wvst.fig}}
\end{center}
\end{figure}

\begin{figure}
\begin{center}
\plotone{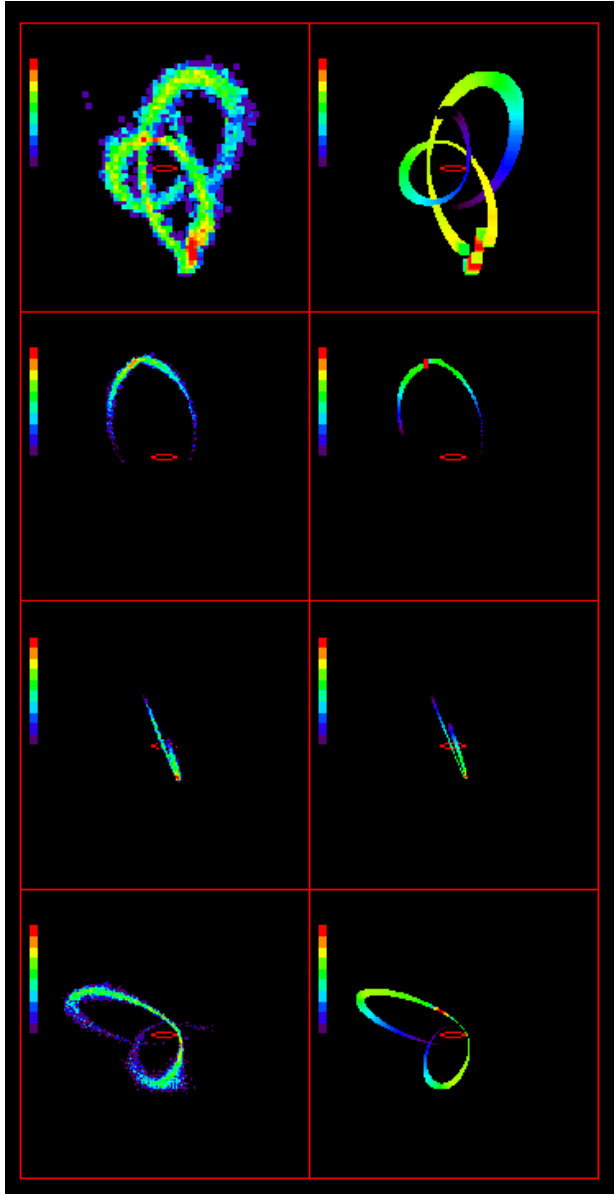}
\caption{
The left-hand panels show the surface brightness calculated
from final positions of particles in four
simulations of satellite destruction (see Table \ref{simstab} for
parameters) within a box $400\times400$~kpc around the
parent galaxy. 
In all cases the parent galaxy is lying edge on 
along the $X$-axis (indicated by the red ellipse)
and the satellite orbits
are viewed at random orientations.
The color bar runs
from $33-28$~mag/arcsec$^2$ in blocks of 0.5 mag in the R band,
assuming a mass-to-light ratio $\Upsilon = 10$. 
(Note that only the yellow and red portions of these streamers would be 
detectable with current techniques.)
The right-hand panels show the corresponding predictions
for the simulation results from semi-analytic methods in which the
satellite mass, orbit, mass-loss rate and mass-to-light ratio are
supplied as input.
\label{surf.fig}}
\end{center}
\end{figure}

Figures~\ref{wvsr.fig} and \ref{wvst.fig} demonstrate that
debris populations formed within the last few 
pericentric passages of a satellite in either spherical or
mildly oblate potentials will have very similar
width and heights that are not much greater than those set
at disruption.
Throughout the remainder of this work, we will assume
that $w\sim h\sim w_0$ for $t<4T_R\sim 3T_\Psi$
is adequate for our purposes --- from Figure \ref{wvst.fig} we estimate
that this approximation is correct to within a factor of two either
up or down (or within one magnitude in surface brightness) for all
the times considered. 
Within this time we also expect debris streams to remain distinct in
coordinate space for satellites with masses less than about
3\% of the mass of their parent Galaxy that is enclosed within the
pericenters of their orbits (or tidal factors less than $s_{\rm lim} =0.31$).
We adopt this as a reasonable limit since we were able to
predict (to within a factor of two) the width of the streamers  in
a simulation with $s=0.15$ for $>10 T_R$ (red points in Figure \ref{wvst.fig}).
This translates to an upper limit $m_{\rm sat}^{lim}$
on the mass of the satellite that can be
considered at a given radius. For example, for a satellite on an orbit
with pericenter $R_p$ orbiting in a galaxy with circular velocity
$v_{\rm circ}$, this limit is
\begin{equation}
	m_{\rm sat}^{lim}=\left({s_{\rm lim} \over 0.31}\right)^3
\left({v_{\rm circ} \over 200 {\rm km/s}}\right)^2
\left({R_p \over 10 {\rm kpc}}\right) 3\times 10^{9} M_\odot. 
\end{equation}

The above 
assumptions appear contrary to the findings of Helmi \& White (1999)
because they examined debris populations of ages $t>>4T_R$,
with the aim of understanding the structure of the velocity
distribution of debris stars.
In fact, our estimates for the surface brightness of
features in \S 3 (which are upper limits to the true surface
brightness because of the above assumptions) 
suggest that even the best photometric observations of external
galaxies today 
are not capable of detecting features much older than
$t=4T_R$, either because they are too
faint, or because of confusion by disk or halo light.

\section{Observability}

In external galaxies, debris trails generally will not be resolvable   
into individual stars.  Instead, they will be recognized in deep 
wide-field photometry as elongated regions of enhanced surface 
brightness around a primary galaxy, perhaps still associated with 
a disintegrating satellite. In this section, we present
a simple estimate for the surface brightness of a debris trail 
generated by a satellite of given properties orbiting in a known 
spherical potential.  This approximate form is then used to 
assess the observability of trails as a function of their age, 
the mass of the satellite, and that of its parent body.  
At the end of this section, sources of confusing background are 
compared to the surface brightness of observable trails. 

\subsection{Surface Brightness of Debris Trails}

Consider a satellite of mass $m$ and constant mass-to-light ratio $\Upsilon$
on a circular orbit at galactocentric radius $R$ around a galaxy with 
circular velocity $v_{\rm circ}$.
If the satellite loses a fraction $f$ of its mass 
in the time $t$ that it takes the debris to spread through an 
angle $\Psi$ then, using equations (\ref{psi}) and (\ref{width}),
the debris is expected to have an apparent surface brightness 
(in mag/arcsec$^{2}$) given by 
\begin{eqnarray}
\label{mu}
	\mu_\nu(t)&=&-2.5\, {\rm log}\left[ {L_{\nu} \over  L_{\odot, \nu}} 
		\left({(10\, {\rm pc})^2 \over w R\Psi}\right)
		\left(1\, {\rm arcsec} \over 1\, {\rm radian}\right)^2
		\right]+M_{\odot, \nu} \nonumber \\
	   &=& -2.5\, {\rm log}\left[ f 
                 \left({10 M_{\odot}/L_{\odot, \nu}\over \Upsilon}\right)
		{\left(1\, {\rm Gyr} \over t\right)}\right] \nonumber \\
	   &&  -{2.5 \over 3}\, {\rm log}\left[	
		{\left(v_{\rm circ} \over 200\, {\rm km/s}\right)}
		{\left(m \over 10^8 M_\odot\right)}
		{\left(10\, {\rm kpc} \over R\right)}\right] + 
		23.9+M_{\odot, \nu}~,
\end{eqnarray}
where $M_{\odot, \nu}$ is the absolute magnitude of the Sun in the waveband
of interest and $L_\nu=(f \, m/\Upsilon) L_{\odot, \nu}$ is the total 
luminosity of the debris trail.  In this approximate representation, 
$w = h$ so that the 
surface brightness is independent of the inclination $i$ of the orbital 
plane to the sky plane. (This simplification is discussed in \S 2.2.)

\begin{figure}
\begin{center}
\plotone{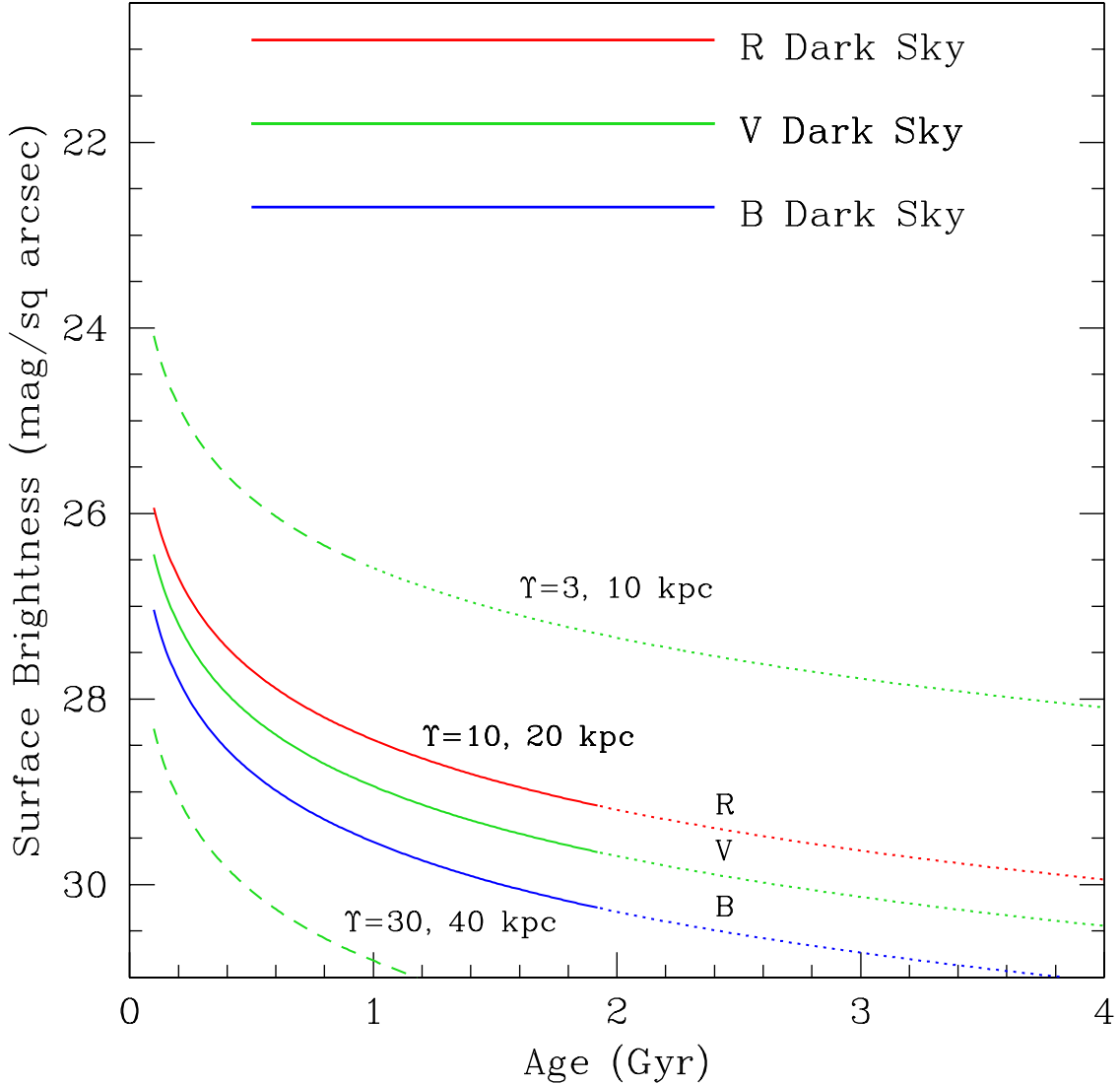}
\caption{Expected surface brightness of tidal debris as a function 
of the age of the trail.  Typical colors of Local Group
dwarfs have been used.  Solid lines indicate the surface
brightness in $BVR$ expected for $10^{8} M_{\odot}$ satellites with
mass-to-light $\Upsilon=10$
that have lost all their mass on circular 20~kpc orbits around primary 
galaxies with $v_{\rm circ}=200$~km/s.  Also indicated for the 
$V$ band are two more extreme cases:
a massive ($9 \times 10^8 M_{\odot}$) satellite with 
low $\Upsilon$ and tight orbit (upper dashed), and a small
($3 \times 10^7 M_{\odot}$) satellite with 
high $\Upsilon$ in a wide orbit (lower dashed.)  
The approximations used here are invalid for
sufficiently old (dotted) streamers.  
For comparison, 
the surface brightness of a moonless night sky is also shown.
Note that due to the typical colors of dwarfs and
the dark sky, surface brightness limits for detection may be
about $0.5-1.0$~mag fainter in $V$ than in $R$.
\label{muvst.fig}}
\end{center}
\end{figure}

Figure~\ref{muvst.fig} shows the temporal decay of surface brightness
estimated from equation~(\ref{mu}) resulting from
the total destruction ($f=1$) of satellites 
similar to the dwarf galaxies in the Local Group.  We have used 
a recent review of Local Group dwarfs (\cite{m98}) as a 
guide to typical luminosities, colors and mass-to-light 
ratios, and have assumed that the primary has $v_{\rm circ} = 200$~km/s,
similar to that of the Milky Way.  Shown as solid lines are the 
surface brightnesses in the $B$, $V$, and $R$ photometric bands for
a $10^8 M_\odot$ satellite with $\Upsilon = 10$ pulled into a debris 
trail along its 20~kpc circular orbit.  Such fiducial structural
parameters 
are roughly compatible with the IC~1613 and Pegasus dwarf  
galaxies in the Local Group.  
The dotted extensions to these lines indicate when a streamer becomes
too old dynamically
($t \ge 3T_\Psi \sim 3\times 2 \pi R/v_{\rm circ}$) 
to allow the direct use of the analytical formulae in this paper, and these
should be regarded as upper limits to the surface brightness.

The detection of such debris
is not trivial, requiring surface brightness photometry reaching 
about 8~magnitudes below sky levels in these bands.
At redder wavelengths, the sky brightness 
presents even more noise compared to the signal expected from dwarfs.  
The $B$-band luminosity and $B-V$ color of dwarfs varies widely depending
on the presence of recent star formation.  On balance, the 
$V$ or $R$ passbands may be the most reliable for the detection of 
tidal streamers in external galaxies.
Satellites that are more luminous than our fiducial value
($L = 10^7 L_\odot$) either due to their larger mass,
smaller $\Upsilon$, or both, and those satellites 
disrupted at smaller galactocentric radii 
($R < 20~$kpc) will generate brighter and thus more easily 
detectable trails.  The opposite is true for less luminous satellites 
and those destroyed at larger radius.  Two such extreme cases 
are shown for the $V$ passband as dashed lines in Figure~\ref{muvst.fig}.
The brighter of these corresponds to the destruction of a 
$L = 3 \times 10^8 L_\odot$, $\Upsilon=3$ dwarf similar to NGC~205 or 
M32 at 10~kpc from its parent galaxy.  A satellite more like 
Phoenix or Carina, with $L = 3 \times 10^6 L_\odot$ and $\Upsilon=30$ 
destroyed at 40~kpc would be too faint for detection with current 
techniques.

Strictly speaking, equation~(\ref{mu}) is only valid for
circular orbits; semi-analytic methods (\cite{j98}) or N-body simulations
can be used to predict variation in surface density along the 
length $\Psi$ for the case of eccentric orbits.  
In the following sections we use this approximation as a guide
to the observability of dynamically young 
features around galaxies and in clusters.  
Debris generated more than about 4 Gyr ago would be expected to have 
surface densities too faint for detection with current capabilities; 
in any event, equation~(\ref{mu}) is not suitable for debris older 
than a few radial periods since precessional effects would then 
cause it to fade faster than $t^{-1}$ (\cite{hw99}). 

\subsection{Astronomical Backgrounds and Target Selection}

In the previous section, simple scaling arguments were used to 
show that if surface brightness levels $\sim$8~magnitudes
below sky can be measured reliably, young tidal debris
from satellites in near orbits around external galaxies can be 
detected (Fig.~\ref{muvst.fig}).  
With extreme care in observational technique and data reduction, 
such ultra-deep surface brightness photometry of galaxies 
is now being performed, generally for the purpose of studying normal, 
but faint, galactic components such as thick disks and stellar halos 
(\cite{mbh94}; \cite{fry99}; \cite{matthews99}; \cite{neeser+00};  
\cite{dalcanton+00}, and references therein), or intracluster light 
(\cite{cal+00}, and references therein).  The very best of these
have achieved detections (over large areas) at
$R \approx$~29~mag~arcsec$^{-2}$.
Only a handful of systems have been studied with sufficient
resolution, depth and sky coverage to detect faint debris trails.

Streamers will be easier to interpret in terms of
dwarf galaxy accretion if the surface brightness can be quantified
and the parent's morphology is largely undisturbed
(suggesting that the dwarf's effect on the
parent --- and hence on evolution of its own orbit --- can be ignored).
A natural sample to survey would be thin, edge-on disk galaxies in which
the orientation is most favorable for detection of low-surface-brightness
features and the
existence of a well-behaved disk suggests that the parent is not significantly
perturbed.  Searches in dynamically active regions
such as galaxy clusters may also yield interesting results, but must
to be interpreted cautiously since material stripped from the primary
galaxies and arcs associated with cluster-lensed background galaxies may
be misinterpreted as debris trails from infalling material.

\begin{figure}
\begin{center}
\plotone{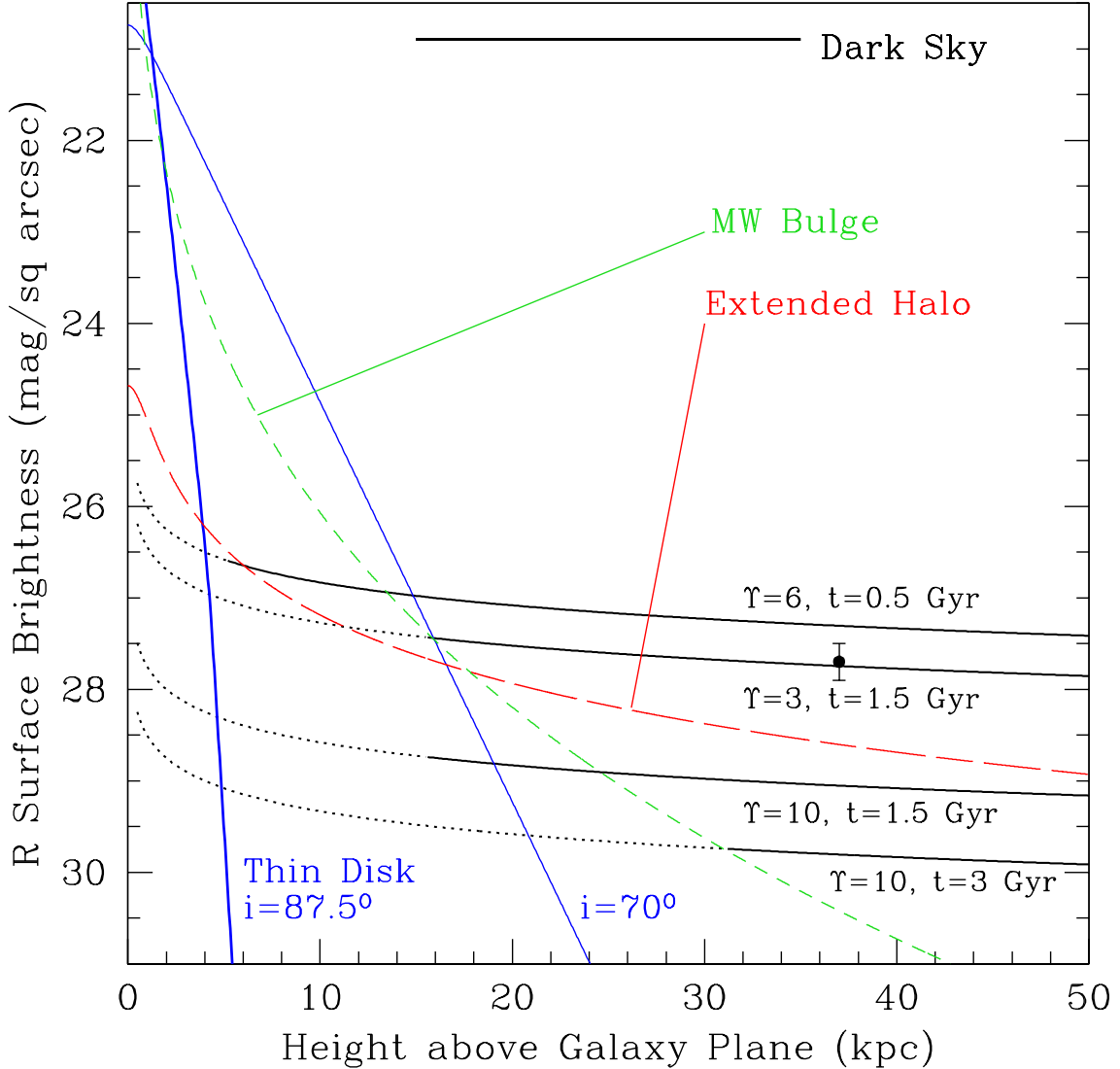}
\caption{
Expected R-band surface brightness of debris trails resulting from 
the destruction of a $10^{8} \, M_\odot$ satellite on a circular orbit
around a thin spiral galaxy with $v_{\rm circ} = 205\,$km/s, as a function 
of the orbital radius (height reached above the galaxy plane).
A variety of streamer mass-to-light ratios $\Upsilon$ and ages $t$ are 
displayed (thin, black line).  
Regions for which the approximations used here (eq.~\ref{mu}) are likely to be
invalid are indicated by dotted, black lines.
For comparison, possible confusing sources of light from the 
galaxy itself are shown: a thin stellar disk inclined by angle $i$ 
with respect to the sky (blue, solid), a $R^{1/4}$ bulge such as
that of the Milky Way (green, short-dash), and a fainter, more
extended $r^{-2}$ halo (red, long-dash).   The disk parameters 
($i=87.5^{\rm o}$), faint extended halo, and $\Upsilon = 3$, $t = 1.5\,$Gyr
streamer are appropriate to the observed and inferred parameters for 
thin edge-on spiral NGC~5907 at a distance of 14~Mpc.  The data point
is the measurement of Zheng et al.\ (1999).
The surface brightness of a moonless R-band night sky is also shown.
\label{muvsr.fig}}
\end{center}
\end{figure}

Measurements of faint surface brightness enhancements near galaxies 
or in clusters are also challenging because in order detect extended
features 8~mag fainter than sky at a $3 \sigma$ level,
all sources of statistical and systematic uncertainty that 
scale with sky brightness must be controlled to within 2 parts in 10$^{5}$.
In addition to the technical demands that this
places on the required number of photons, the precision of 
detector calibration and the background (sky) level determination, 
the observer must also contend with astronomical sources of uncertainty 
due to confusing light from the galaxy itself and its environment. 
The behavior of the most common or troublesome
astronomical sources of confusion is shown in Figure~\ref{muvsr.fig},
together with the estimated surface brightness (from eq.~\ref{mu}) 
of streamers in circular orbits of a variety of radii around a 
typical disk galaxy.   
As in Figure~\ref{muvst.fig},
the dotted lines in Figure~\ref{muvsr.fig} trace 
features that are too old dynamically for interpretation using
the analytical formulae in this paper;
many of these features would also pass too near the disk plane
of the primary to generate circular orbits.

Because the surface brightness of even young debris is so faint,
galaxies that are viewed edge-on
offer the best targets for streamer searches.  We have chosen 
the edge-on Sc galaxy NGC~5907 to illustrate these effects in 
Figure~\ref{muvsr.fig}, and will analyze a streamer around this 
galaxy in a later section.  NGC~5907 is nearly bulgeless and is 
thin, with a scale length to scale height ratio of $h_R/h_z \approx 11$ 
(\cite{mbh94}). The surface brightness profile of its thin disk, 
which is inclined at about $i = 87.5^{\rm o}$ to the sky plane, is shown
as a function of height above the galaxy along its minor axis.  
The thin disk light of similar spirals will swamp that 
of reasonably bright debris streamers --- even at heights of $\sim$20~kpc 
or more --- if the galactic disk is inclined by $i \le 70^{o}$.   

Bulges, like that of Milky Way, are a source of unwanted 
of background light regardless of inclination angle, obscuring all 
but the very brightest debris within 20~kpc and a confusing factor 
out to 30--40~kpc for many streamers of reasonable age, $\Upsilon$,
and total mass.  Even much fainter spheroidal galactic components
can complicate the detection of faint tidal debris if the light 
is sufficiently extended.  Typical stellar halos of spirals have
luminosity densities that decline with galactocentric radii like $r^{-3.5}$ 
or steeper (\cite{saha85}; \cite{zinn85}; \cite{reitzel98}); these 
will provide a confusing background of light only at the smallest 
galactocentric radii.   
Some giant ellipticals and CD galaxies, on the other hand, 
have halo luminosities and globular cluster systems that fall less 
steeply, as $\sim r^{-2.3}$  
{(\cite{harris86}; \cite{bridges91}; \cite{harris95}; \cite{graham96}). 
Such extended faint stellar halos may interfere with the detection 
of faint debris streamers. Indeed the surface brightness of the extended
$r^{-2}$ halo reported in NGC~5907 inward of 10~kpc (\cite{smhb94}; 
\cite{lequeux96}; \cite{rudy97}; \cite{jc98}) is substantially brighter than 
the average surface brightness of its streamer seen at 
larger radii (\cite{s+98}; \cite{z+99}). 

\section{Applications}

\subsection{Interpreting the Remains of Disrupted Satellites}

The intrinsic geometry of a streamer from a totally 
disrupted satellite can be used to estimate the mass $m$ and age $t$ 
of a young streamer, provided that the form of the galactic potential 
can be estimated.  For example, rearrangement of equations 
(\ref{tscale}), (\ref{psi}), and (\ref{width}) for a logarithmic 
potential yields
\begin{equation}
\label{mass}
	m = s^3 M(R_p) = \left({w \over R}\right)^3 M(R_p)
	  \sim 10^{11} \left({w \over R}\right)^3
	  \left({R_p \over 10\, {\rm kpc}}\right)
	  \left({v_{\rm circ} \over {\rm 200\, km/s}}\right)^2 M_\odot ~,  
\end{equation}
and hence the mass-to-light ratio $\Upsilon$ of the satellite, and 
the time since its disruption 
\begin{equation}
\label{time}
        t\sim 0.01 \, \Psi \left({R \over w}\right)
          \left({R_{\rm circ} \over 10\, {\rm kpc}}\right)
          \left({ 200\, {\rm km/s} \over v_{\rm circ} }\right)
          {\rm Gyr}~,
\end{equation}
where $w$ is the width of the streamer at radius $R$, $\Psi$ is 
its angular length, and $R_p$ is the pericentric distance of the orbit.
In the second equation we have replaced $T_{\Psi}$ by 
$2 \pi R_{\rm circ}/v_{\rm circ}$ where $R_{\rm circ}$ is the radius of the
circular orbit with the same energy as the true orbit. (Recall that
the time periods of orbits depend primarily on their energy.)
Of course, we cannot
measure $R_{\rm circ}$ directly, but can approximate it as being
halfway between the adopted apocenter and pericenter\footnote{
For orbits with $R_p/R_a>0.1$ in a logarithmic potential,
$0.9 R_{\rm circ}<(R_p+R_a)/2<R_{\rm circ}$.}.
Thus, for a logarithmic potential, the only information about 
the primary galaxy that one requires is an estimate of the circular 
rotation speed, which can be directly obtained from a measured rotation curve 
or the Tully-Fisher relation.   The estimates of $m$ and $t$ depend 
only linearly on the cosmological distance to the galaxy.  

More troublesome is the ambiguity 
introduced by the (generally) unknown inclination $i$ of the streamer 
to the sky plane at various positions along the orbit.   
In order to simplify the discussion, we assume that the
streamer is co-planar, which is reasonable for the age of
observable features.
We define an orthogonal coordinate system with the 
$X$ and $Y$-axes in the plane of the sky  
and the $Z$-axis perpendicular to these along the line of sight.
If the $Y$-axis lies along the intersection of the sky and
orbital planes, then $R=R^{\rm obs}$ (where $R$ and $R^{\rm obs}$ 
are the true and observed distances from the center of the parent
galaxy) along this axis, and $R=R^{\rm obs}/\cos i$ along the $X$-axis.

If the observed morphology of the streamer indicates that its orbit
is nearly closed, we can assume that the orbit is
nearly circular and $i$ can be estimated from the
ratio of the observed semi-major $Y=a$ and semi-minor axes $X=b$ of the orbit,
namely $\cos i = b/a$.
In general, however, the orbit will not be closed
and only a partial arc may be ``illuminated'' by detectable debris. 
In this case, the change of the width of the streamer along the arc 
can provide a useful diagnostic.  Since we have assumed that 
$w = h$ at any point along the young streamer, projection has no effect on 
the observed width of the streamer, which does, however, increase with 
$R$ in any reasonable potential.  The observed width $w$ of the streamer 
then becomes an indicator of relative galactocentric distance $R$
(with $R_2 \approx R_1 (w_2/w_1)$ for any two points 1, 2 along the orbit
in a logarithmic potential).
For example, in the top panels of Figure~\ref{surf.fig} the orbit
is face-on and the width of the
streamers increase with projected distance from the galaxy,
while in the third panels the orbit is nearly edge-on
yielding a streamer with very different
widths at similar projected radii.

As a final check, semi-analytic models or N-body simulations can be 
used to check the consistency of the measured surface brightness $\mu$ 
and width $w$ at several points along the feature for given model 
assumptions.  An example of applying these simple ideas is
presented in the following subsection.

\subsubsection{Worked Example: NGC 5907}

Shang et al.\ (1998) report the discovery of a partial ``ring'' of 
light around the edge-on disk galaxy NGC~5907. The arc is observed
to be brightest and widest in the northeast half, where the 
average surface brightness is $28.0 \pm 0.2$~mag~arcsec$^{-2}$ in 
their intermediate band filter centered on 6660 \AA, corresponding 
to about 27.7~mag~arcsec$^{-2}$ in broadband R (\cite{z+99}). 
The inferred broadband $R-I$ color is about $0.5 \pm 0.3$, consistent
with Local group dwarf satellites (\cite{m98}). 
No gas has been detected in H$\alpha$ or HI emission lines associated 
with the arc of light.
Shang et al.\ (1998) interpret the arc as debris from the destruction of a 
satellite on an elliptical orbit around the galaxy. 

We adopt the satellite debris interpretation, but instead assume 
that the material is on a rosette rather than elliptical orbit, as 
is consistent with the assumption that NGC~5907 is embedded in a massive 
dark halo with circular velocity $v_{\rm circ}=205\,$km/s (deduced from
VLA observations by \cite{s+98}).  Furthermore, given the uncertainty
introduced by the light from the galactic disk and obvious confusion  
from foreground stars along some portions assumed trajectory 
(the Shang et al.\ data were taken in 4 -- 6\arcsec\ seeing),
we analyze only the clear northeast portion of the streamer, 
which subtends $\Psi \approx 2 \pi/3$ radians of arc.

Assuming a distance to NGC~5907 of 14~Mpc (\cite{zepf+00})  
and the parameters given in Shang~et~al.~(1998) and Zheng~et~al.~(1999), 
the widest portion of the streamer has $w\sim 4.6\,$kpc 
and lies at a projected distance of $R^{\rm obs} = 49\,$kpc from 
the center of the galaxy.  The narrowest portion that can be  
observed cleanly lies just above the projected southeastern 
intersection of the streamer and galaxy at $R^{\rm obs} \approx 18\,$kpc, 
where the width is $w\sim 2.3\,$kpc.  This latter position may or may 
not be the pericenter of the orbit, and is displaced by about $\pi/2$ 
from the widest portion of the streamer.  

Since the width clearly increases 
with projected distance from the galaxy, it is reasonable to assume 
that the satellite's orbit is not highly inclined with sky plane.  
The linear relationship between $w$ and $R$ in an isothermal potential 
suggests that the ratio of the true galactocentric distances of the
widest and narrowest portions of the observed streamer is nearly equal
to the ratio of their true widths.
If the apparent 
major axis of the orbit lies near the sky plane, an inclination 
$i \sim 40^{\rm o}$ is implied.  In this case, $R \approx R^{\rm obs} = 49\,$kpc
for the widest portion of the arc, 
whereas $R \approx R^{\rm obs} / \cos i = 24\,$ kpc for the narrowest 
portion of the arc.
Since tidal streamers lie inside or outside their satellite's orbit
depending on whether they are leading or trailing the satellite,
we assume for the purposes of this illustration
that $R_p=24$~kpc and $R_{\rm circ}=37~$kpc.
 
Inserting these values in equations (\ref{mass}) and (\ref{time}),
we estimate that the streamer comes from the destruction of
a $m \approx 2 \times 10^8M_{\odot}$ satellite $t \approx 0.8\,$Gyr ago. 
The mass-to-light ratio 
$\Upsilon$ of the progenitor can be estimated by noting that 
Shang~et~al.\ (1998) give a lower limit of $14.7 \pm 3$ for the total
apparent magnitude in the 6660\,\AA\ band, assuming that the streamer has 
its average detectable surface brightness throughout $2 \pi$ radians.  
Converting to broadband R and considering only the $2 \pi/3$ easily-detected  
arc, we obtain an estimate of $ \le 7 \times 10^{7}\, L_{R, \odot}$ 
for the observed debris, and thus estimate that the debris trail has  
$\Upsilon_R \ge 3 M_{\odot}/L_{R, \odot}$, or about 
$\Upsilon_V \ge 4 M_{\odot}/L_{V, \odot}$.  
Given the uncertainties in the distance to NGC~5907 
(\cite{zepf+00}) and in the apparent magnitude of the debris, these
mass-to-light ratios are uncertain by nearly a factor of two. 
These inferred satellite parameters are consistent with a progenitor
similar to dwarf satellites such as NGC~185 or Pegasus in the Local 
Group, and the Fornax dSph (\cite{m98}).

Because much of the angular length $\Psi$ may go undetected
due to surface brightness constraints, the streamer's age as estimated above
is a lower limit on the time since satellite disruption.
Indeed, the deep photometry of Shang et al.\ (1998)
hint at a feature on the southwestern side of the galaxy, plausibly 
connected to the clearer arc on the other side, but making a sharp 
turn as the features cross the disk of NGC~5907 (in projection).  
If we interpret this feature as real and part of the debris trail, 
then $\Psi \ge 2\,\pi$ and the inferred age of the entire streamer 
increases to 2.4~Gyr.  
As a final check, the observed
surface brightness of the ring (27.7~mag~arcsec$^{-2}$ in broadband R,
\cite{s+98}; \cite{z+99}) agrees with
our analytical estimates for debris from
a $2\times 10^8 M_\odot, \Upsilon=3$ satellite
on a circular orbit at $R=37$~kpc with age in the range $0.8-2.4$ Gyrs
(see Fig.~\ref{muvsr.fig}).

We do not claim that these estimates are unique, particularly as they 
depend on the precise geometry of the streamer.  Our estimates are, 
however, compatible with those of Reshetnikov and Sotnikova (2000),  
who conclude that the stellar arc around NGC~5907 has 
$\Upsilon_B \ge 3 - 7 M_{\odot}/L_{B, \odot}$ 
and an age $t \le 1.5\, $Gyr 
based on recent N-body simulations of the disruption of a dwarf companion 
moving in the polar plane of the NGC~5907 gravitational potential.
Our independent semi-analytic simulations of this system, performed without 
knowledge of the Reshetnikov and Sotnikova work, confirm that the 
faint southwestern feature could plausibly be part of the same 
debris trail (see their Fig.~4).

\subsection{Limits on the Mass Loss Rate from Observed Satellites}

If we know that a satellite (with a dynamically-estimated mass)
is at a projected distance from its parent (with measured $v_{\rm circ}$)
but has no associated tidal tail to some limiting surface brightness
$\mu_{\rm lim}$,
we can place an upper limit on its fractional mass loss rate by re-arranging
equation (\ref{mu}):
\begin{equation}
	{df \over dt} < \left({\Upsilon \over 10 M_\odot/L_\odot}\right)
\left[\left({200 {\rm km/s} \over v_{\rm circ}}\right)
\left({10^8 M_\odot \over m_{\rm sat}}\right)
\left({R_p \over 10{\rm kpc}}\right)\right]^{1/3}
10^{{(M_{\odot,\nu}+23.9-\mu_{\nu,{\rm lim}}) / 2.5}}\,\, {\rm Gyr^{-1}}.
\label{dfdt}
\end{equation}
An example is given below.

\subsubsection{Worked Example: Satellites of M31}

\begin{table}
\begin{center}
\begin{tabular}{ccccc}
name	& $m_{\rm sat}$ & $R^{\rm obs}$  & $(M/L)_V$       & $df/dt$ \\
	& $M_\odot$	& kpc            &                 & Gyr$^{-1}$ \\
\hline
M32	& $2.1\times 10^9$ & 5.4	& 5.6 & $<$ 1.4 \\
NGC~205  & $7.4\times 10^8$ & 8.2	& 2.0 & $<$ 0.85 \\
\end{tabular}
\end{center}
\caption{Properties of M31 satellites:
the satellite mass, $m_{\rm sat}$;
projected distance between the
satellite and M31, $R^{\rm obs}$; mass-to-light ratio in the $V$ band;
and upper limit on the fractional mass loss rate, $df/dt$.
Observed quantities are from Mateo (1998).
\label{satstab}
}
\end{table}

Choi \& Guhathakurta (2001) have recently completed a CCD mosaic of M31
which allows them to measure the outer isophotes of M32 and NGC~205 to
a limiting surface brightness of $\mu_{B}=$27 mag/arcsec$^2$.
Although their survey does reveal interesting features at low surface
brightness levels close to each
satellite, they are unable to trace these features beyond more
than a couple of tidal radii.
Adopting a value of $v_{\rm circ}=235\,$km/s (\cite{braun91}) 
for M31's dark halo
we can calculate an upper limit to $df/dt$ for each satellite using their
dynamical masses, observed $M/L$ and projected
distances from M31 (see Table \ref{satstab}).
This example is intended only as an illustration;
at this limiting surface brightness level, the limits on $df/dt$ are very
weak.  In these particular cases, a stronger limit may be
found by looking at the features close to each satellite
(see \cite{jcg01}).

\subsection{Tidal Streamers and Galaxy Formation}

Can cosmological models of     galaxy  formation be tested  using    a
sensitive survey for debris trails?

N-body simulations  naturally   incorporate   the  physics    of tidal
disruption  and debris dispersal.   With  additional assumptions about
the relationship between the  formation of stars  and the dark halo in
which they  reside, the frequency  of  low surface brightness features
observable around galaxies   today could be  predicted  within a given
cosmological model.  This approach is limited by the resolution of the
simulations to placing  only lower limits on the  number of trails  we
might expect  to  observe.  First,  as with identifying  halos, there
will be a mass  limit below which a  stream from a disrupted halo will
not be resolved  (presumably at higher  mass than the satellite itself
since the particles   represent lower   surface brightness  features).
Second, the finite resolution will cause random potential fluctuations
that  will  scatter particles    (i.e.,   numerical relaxation  --  see
\cite{w93} for a  general discussion) and  tend to artificially reduce
the surface brightness  of debris streams.
For example, Johnston, Spergel \& Haydn (2001)
found that a  minimum of  $10^7$ particles  were needed in  the parent
Milky Way-like halo in order for a  trail from a $10^8-10^9 M_{\odot}$
satellite following an Sgr-like orbit to maintain its coherence over
4~Gyrs.  This finding is clearly dependent on the  mass and orbit of the
satellite and   the  mass  distribution  in   the parent galaxy,   but
nevertheless indicates the scale of  the problem.  It suggests that an
appropriate   N-body  approach  would be  to  simulate  $\sim$100 such
galaxies, each with $10^7$ particles within a cosmological context and
then   analyze them to  assess  the number of   trails  expected to be
observed around a similar sample of real galaxies.

One  way of testing whether  such a computationally expensive study is
warranted  is to  model  satellite accretion  history
in  a semi-analytic  fashion that allows multiple
realizations to be   performed  rather quickly, thus  providing  ample
statistics and the  ability to probe a  large parameter space of input
assumptions (e.g. \cite{cwg93}; \cite{c+94}; \cite{sp99}).
Current models of this
kind use  Monte-Carlo methods  based  on the  extended Press-Schechter
formalism (\cite{kw93}; \cite{lc93}; \cite{sk99}),
and have proven successful
at   reproducing the merging  histories of  dark  halos seen in N-body
simulations (\cite{lc94}; \cite{s+00}).  As    with  N-body  simulations,
associating galaxies  and stars with dark halos  must be  done using
some  prescription.  In addition, for application to satellite
debris observations, some assumptions
about the  orbital distributions  and   the tidal disruption  of   the
accreted  objects  must  be made.    The morphology   and  surface
brightness of  the streamers can then  be predicted using methods
presented here and in Johnston (1998).
Unlike the  N-body  case, this approach,  although
less exact, suffers from no resolution limits.

\subsubsection{Worked Example: Milky Way Satellite Debris in $\Lambda$\rm CDM}}

As an illustration of this idea, we have used the model of
Bullock, Kravtsov \& Weinberg (2001)
to  provide  the  accretion  histories  of
an ensemble  of 100  halos, each with final $v_{\rm  circ} =
220$~km~s$^{-1}$, which are assumed to host
Milky  Way-type galaxies.  The model assumes a  $\Lambda$CDM cosmology with
$\Omega_m = 0.3$, $\Omega_{\Lambda} = 0.7$, $h=0.7$,and $\sigma_8=1.0$,
and provides
masses, approximate disruption  times, and orbital evolution  for each
disrupted satellite.
We  estimate the mass-to-light ratios  of  the disrupted satellites by
applying the same hypothesis used by Bullock, Kravtsov \& Weinberg (2000)
to match the observed abundance of  Local Group satellites today: that low-mass
satellites with  virial  temperatures below  $\sim 10^{4}$K ($v_{circ}
\la 30$km s$^{-1}$) can  only accrete  gas   before the universe was
reionized\footnote{
 Without the reionization solution,  the number  of surviving observable
 satellite galaxies in the Milky  Way today would  be over-predicted by a
 factor  of $\sim 10$.} at  $z=z_{re}$.
Under this assumption, the  final baryonic mass of a  satellite halo
of  mass $m$
will  be $f_m m  (\Omega_b/\Omega_m) \simeq  0.14 f_m m$.   If $v_{circ} <
30$~km s$^{-1}$, $f_m=f_{re}$ represents the  fraction of the halo's mass
that was in  place at $z=z_{re}$.   (We assume $z_{re}=8$.)  For halos
with sufficiently deep  potential wells  ($v_{circ} > 50$~km  s$^{-1}$)
the photoionizing background will not affect gas accretion, and $f_m=1.$
For intermediate masses, we assume  that $f_m$ varies linearly
between  $f_m=f_{re}$ and $1$ as the
halo size varies from $v_{circ} = 30$  to $50$~km s$^{-1}$ (\cite{tw96}).
Finally,  if a  fraction $\epsilon_*  =0.5$ of  this baryonic  mass is
converted to a stellar  population  with current mass-to-light ratio
$M_*/L_V=2$, this implies $\Upsilon_V  =  30 f_m^{-1}$.  With $\Upsilon_V$  and
the  other disrupted satellite  properties provided  by  the model, we
then use equation~(\ref{mu}) to estimate   the surface brightness  of
features today.

\begin{figure}
\begin{center}
\plotone{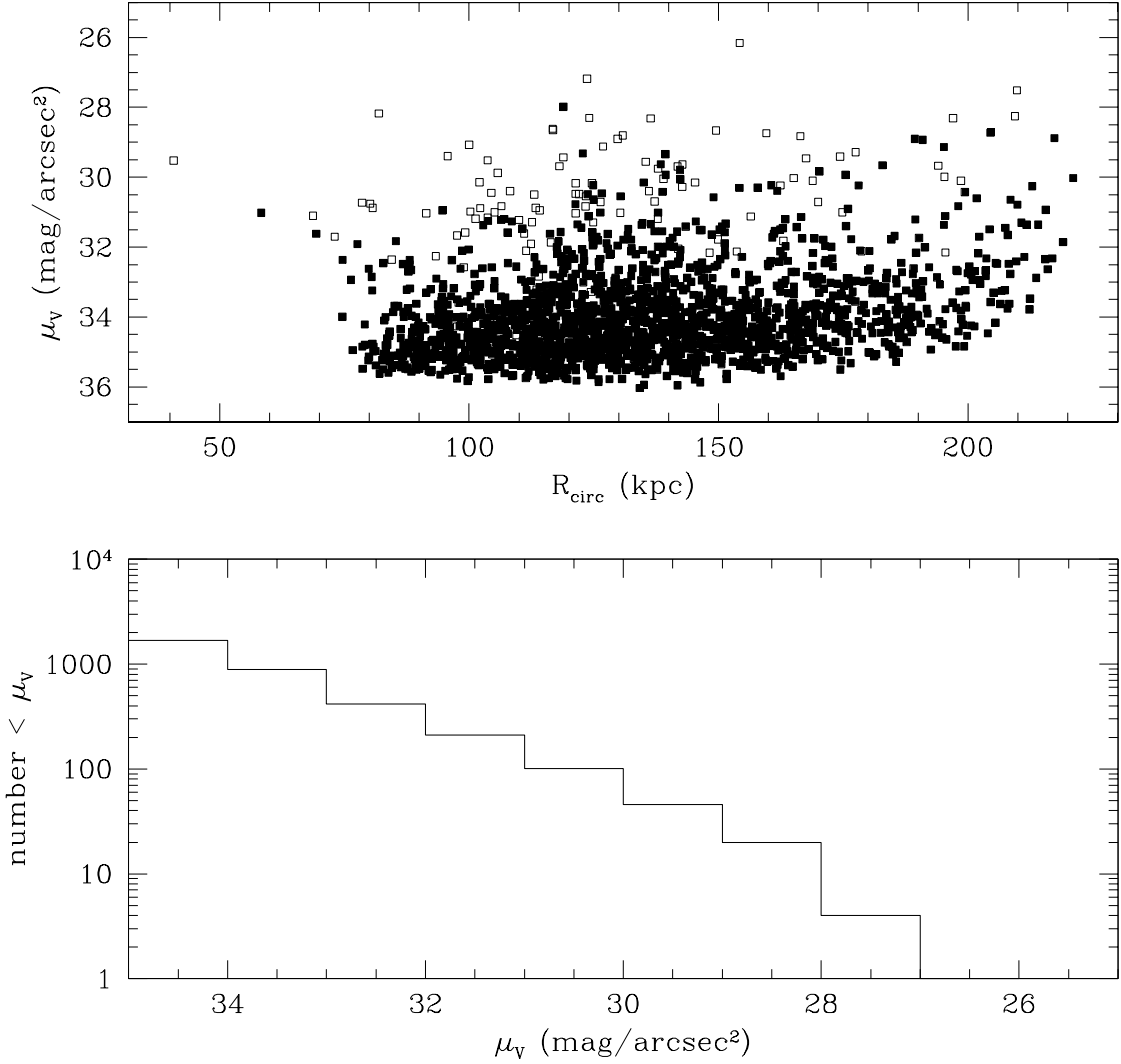}
\caption{
The top panel shows the position of debris in the $\mu_V-R_{\rm circ}$ plane in
100 semi-analytic realizations of galaxies. Only those points for
disruption events that satisfy $t<3T_\Psi$, and that have occurred
since the parent galaxy has accreted 90\% of 
its mass are plotted. The filled points highlight those for which $s<0.31$.
The bottom panel plots the cumulative number of trails brighter than
a given limiting surface brightness.
\label{cosmo.fig}}
\end{center}
\end{figure}

The closed and open symbols in the top panel of Figure~\ref{cosmo.fig}
indicate
the final positions in the $\mu-R_{\rm  circ}$ plane of all disruption
events in our simulated ensemble  that that 
satisfy our  age criterion, $t < 3 T_\Psi$, and have occurred since each  parent
galaxy has accreted 90\%  of  its mass.  The latter requirement  was
imposed   to ensure that additional orbit   evolution of the debris is
unlikely to  be an important  effect. 
Since the age of accretion events decreases with distance from the
parent galaxy, one consequence of this requirement
is that most of the trails have $R_{\rm circ}>
50$ kpc.
The trend of increasing $R_{\rm circ}$ with decreasing age also accounts for
the fact that trails at large radii have higher 
average surface brightness than those at smaller radii.

Plotting against $R_{\rm circ}$ in Figure~\ref{cosmo.fig}
does not directly illustrate the expected distribution of
debris because most of the satellites were destroyed along
eccentric orbits with $R_p << R_{\rm circ}$.
This is reflected in the distribution of the closed symbols which indicate  the
subset of these events that also satisfy our criterion, $s<0.31$.
The total number of features
brighter than a given surface brightness is shown in
the bottom panel of Figure~\ref{cosmo.fig}, indicating clearly
that,  within   the  chosen  cosmology  and star-formation
prescription, we would expect  to see many  tens of features  brighter
than $\mu_V=30$~mag  arcsec$^{-2}$ (about ten of which
satisfy $s<0.31$ and hence are easiest to interpret) 
in such  a survey of 100 parent
galaxies.   Moreover, these
features tend to lie at large ($>50$~kpc) radii from their  parent galaxy
where they are less likely to be confused by background disk, bulge or
halo light. (Fig.~7 shows that only very extended stellar halos
are brighter than $\mu_R=29$~mag  arcsec$^{-2}$ or $\mu_V=29.5$~mag  
arcsec$^{-2}$ at $R=50$~kpc.)
We  defer to a  future paper a  more
complete discussion  of the dependence  of  satellite debris
characteristics on star formation prescription and cosmology.  In the next
section we discuss the feasibility of such an observational survey.

\section{Observational Expectations for the Future}

The apparent scarcity of observed debris streamers in external 
galaxies need not necessarily imply that they are not present, 
since the arguments we have presented indicate that most are
probably at or below current detection levels.  Special
photographic emulsions may provide an excellent way of searching 
for such features over a wide field (\cite{mh97}), but calibration 
of the surface brightness levels needed for the best modeling 
will likely rely on charge coupled devices (\cite{weil97}).
Since more photons are available for imaging than spectroscopy of
individual lines, detection of low surface brightness features in
external galaxies will always be easier than the measurement of their
line-of-sight velocities.

For ease of detection, one would like:\\
\hglue 1.25cm  $[1]$ clear detection of diffuse light several magnitudes below sky\\
\hglue 1.25cm  $[2]$ simultaneous imaging of the complete streamer and the background sky\\
\hglue 1.25cm  $[3]$ minimal confusion from astronomical backgrounds\\
\hglue 1.25cm  $[4]$ control over systematic uncertainties, eg., flat-fielding and
		scattered light\\
\noindent
The first goal requires a large number of photons, and the second a large detector.
The third goal can be partially addressed with appropriate target selection (\S 3.2),
while the last will require care in instrument design and observational strategy.
In principle, the last goal is likely be the most difficult to achieve, but
the first two are necessary, if not sufficient, conditions.  We thus
focus now on meeting the first and second goals with
current and future instrumentation, namely the Advanced Wide Field Camera (ACS/WFC)
on the 2.4m HST, the FORS imaging system on the 8m VLT, and proposed
super-aperture telescopes, such as the 30 CELT or 100m OWL.

Using NGC~5907 as a guide, we consider the possibility of
detecting a streamer with an observed length of 50~kpc and
width of 5~kpc, to a depth that is one magnitude fainter
than the average surface brightness R = 28~mag~arcsec$^{-2}$
currently measured for the NGC~5907 arc.
A rough upper limit to the physical size of the required field is 400~kpc, which
allows 200~kpc on either side of the center of a galactic potential well for
detection of satellite tails (see Figs.~6 and ~\ref{cosmo.fig}).  
The angular size
of the required field will depend on the distance to the galaxy; a 400~kpc field
corresponds to about 800\arcsec\ at 100~Mpc.  We now ask how long an
integration time would be needed on various instruments in order
to acquire the photons required for a $2 \, \sigma$ detection of an area
of the streamer equal to its typical width squared.  Since a streamer
might be expected to have a length at least 10 times its width (\S 2.2), the
complete streamer would then lie $\sim$$6 \, \sigma$ above the background noise.

Spaced-based observations such as those with the ACS/WFC have the advantage
of low sky background, but also have smaller apertures relative to the most powerful
ground-based imaging instrumentation.  The $\sim$200\arcsec\ field of
the ACS/WFC is well-matched to primary galaxies at 400~Mpc, for which
a total exposure of $\sim$1~hour is required to achieve the desired S/N
with the standard configuration in typical conditions.
The $\sim$400\arcsec\
field size of the FORS imaging system on the VLT is suited to more nearby
galaxies at distances of 200~Mpc,
for which $\sim$12~minutes is required to
reach the fiducial photon noise in typical, dark conditions on Paranal.
A streamer of similar size, but $R$ = 30~mag~arcsec$^{-2}$, would yield the required
photon S/N in 1.5~hours with the VLT.  The increased aperture afforded by
planned telescopes such as CELT or OWL could reduce this time by
factors of 3 to 10.
Exposure times with the ACS could be halved
if the search was resticted to streamers of similar
angular (rather than physical) size as those probed by the VLT.  In
general, however, for the simple comparison made here, larger aperture
wins over the reduced sky background.

We stress that the reliability of very faint detections
will be dominated by systematics rather than photon noise, so that
these estimates should be considered a strict lower limit to the
total time required.  Experience suggests that
at surface brightnesses of $26 < R < $~28~mag~arcsec$^{-2}$,
systematic uncertainties are comparable to photon noise
in their contribution to the error budget,
so that exposure times 2 to 4 times longer than those estimated
from photon statistics
are required to realistically achieve the desired total S/N.
At fainter levels, systematic uncertainties may completely dominate
unless extreme care is taken in observational strategy and
instrument design.
Finally, note that imaging designed
to reliably {\it photometer\/} debris streamers along their length
will require considerably longer ($\times 10$) exposures
than those aimed at detection only.
Nevertheless, these estimates indicate that searches
for young debris trails as faint as $R < $~29~mag~arcsec$^{-2}$
are feasible now around a sample of galaxies
200~Mpc distant from us.  Planned super-aperture telescopes would allow
the study of a fainter and older streamers around a larger sample of
primaries.

\section{Conclusions}

We have developed and tested in this paper
a simple semi-analytic formalism for following the
surface brightness and morphology of debris during the first 
few orbits after a satellite is destroyed.
Using the properties of Local Group satellites and extended light
distributions around external galaxies
we discussed to what extent the parent galaxy's light 
should obscure such features.
We then applied our models to
three representative astronomical applications:
(1) estimation of the mass-to-light ratio $\Upsilon$
and streamer age of the totally disrupted satellite responsible for NGC~5907's
debris arc,
(2) derivation of the upper limit to the mass-loss rate $df /  dt$ of
partially-destroyed satellites such as would be expected for M32 or NGC~205, and
(3) exploration of galaxy formation in a $\Lambda$CDM cosmology.

We find that the debris trail in NGC~5907 is well-modeled by disruption of
a satellite with mass $m \approx 2\times 10^8 M_\odot$ and mass-to-light
ratio $\Upsilon _R = 3$ originally on an orbit of approximate radius
$R=37$~kpc disrupting $0.8-2.4$~Gyrs ago.
We also find that the lack of very bright debris associated with the satellites
of M31 indicates mass-loss rates $df/dt < 1.5$~Gyr$^{-1}$.
Finally, we find that a first examination of
realizations of 100 galaxies in a standard $\Lambda$CDM
cosmology, yields over forty features brighter than $\mu_V=$~30~mag/arcsec$^2$.
Such $V$ surface brightnesses are at the limit of current capabilities, but
--- if systematics can be controlled --- may become routine
with proposed telescopes such as OWL or CELT, allowing surveys for
satellite debris in external galaxies to yield constraints
on hierarchical models of galaxy formation.

\acknowledgements
The authors would like to thank Mike Irwin for helpful comments on 
Sagittarius' Carbon star population.
KVJ and PDS would like to thank the Institue for Advanced Study for 
hospitality during a crucial stage in the development of this study.
KVJ was supported in part by NASA LTSA grant NAG5-9064.
JSB was supported by NASA LTSA grant NAG5-3525 and NSF grant AST-9802568.

\begin{thebibliography}{}

\bibitem[Braun 1991]{braun91}
Braun, R. 1991, \apj, 372, 54

\bibitem[Bridges, Hanes \& Harris 1991]{bridges91}
	Bridges, T.J., Hanes, D.A., Harris, W.E., 1991, AJ, 101, 469

\bibitem[Bullock, Kravtsov \& Weinberg 2000]{b+200}
     Bullock, J.S., Kravtsov, A.K., Weinberg, D.H., 2000, ApJ, 539, 517

\bibitem[Bullock, Kravtsov \& Weinberg 2001]{b+100}
     Bullock, J.S., Kravtsov, A.K., Weinberg, D.H., 2001, ApJ, 548, 33


\bibitem[Cole et al. 1994]{c+94}
         Cole, S., Arag\'{o}n-Salamanca, A., Frenk, C.S., Navarro, J.F.,
          Zepf, S., 1994, MNRAS, 271, 781

\bibitem[Calc{\'a}neo-Rold{\'a}n et al.\ 2000]{cal+00} 
Calc{\'a}neo-Rold{\'a}n, C., Moore, B., Bland-Hawthorn, J., Malin, D.\ and 
Sadler, E.\ M.\ 2000, \mnras, 314, 324 

\bibitem[Choi \& Guhathakurta 2001]{cg01}
Choi, P. \& Guhathakurta, P. 2001, {\it in preparation}


\bibitem[Dalcanton \& Bernstein 2000]{dalcanton+00}
Dalcanton, J.J. \& Bernstein, R. 2000, astro-ph/0005327


\bibitem[Fry et al.\ 1999]{fry99}
Fry, A.M., Morrison, H.L., Harding, P., \& Boroson, T.A. 1999, \aj, 118, 1209

\bibitem[Graham et al.\ 1996]{graham96}
	Graham, A., Lauer, T.R., Colless, M., Postman, M., 1996, ApJ, 465, 534

\bibitem[Harris 1986]{harris86} 
	Harris, W.E., 1986, AJ, 91, 822

\bibitem[Harris, Pritchet \& McClure 1995]{harris95} 
	Harris, W.E., Pritchet, C.J., McClure, R.D., 1995, ApJ, 441, 120

\bibitem[Helmi \& White 1999]{hw99}
Helmi, A. \& White, S.D. 1999, \mnras, 307,495

\bibitem[Helmi, White, de Zeeuw \& Zhao 1999]{h+99} 
Helmi, A., White, S.\ D.\ M., de Zeeuw, P.\ T.\ \& Zhao, H.\ 
1999, \nat, 402, 53 
 
\bibitem[Hernquist \& Ostriker 1992]{ho92} 
Hernquist, L. \& Ostriker, J.  P. 1992, \apj, 386, 375

\bibitem[Ibata et al.\ 2000]{i+00}
Ibata, R., Lewis, G. F., Irwin, M., Totten, E. \&
Quinn, T. 2000, astro-ph/0004011
 
\bibitem[Ivezi{\'c} et al. 2000]{sdss00b} 
Ivezi{\'c}, Zeljko et al.\ 2000, \aj, 120, 963 
 
\bibitem[James \& Casali 1998]{jc98}
	James, P., \& Casali, M. M. 1998, MNRAS, 301, 280

\bibitem[Johnston 1998]{j98}
Johnston, K.V. 1998, \apj, 495, 297

\bibitem[Johnston, Choi \& Guhathakurta 2000]{jcg01} 
Johnston, K. V., Choi, P. \& Guhathakurta, P. 2000, {\it in preparation}

\bibitem[Johnston, Hernquist \& Bolte 1996]{jhb96}
Johnston, K. V., Hernquist, L. \& Bolte, M. 1996, \apj, 465, 278

\bibitem[Johnston, Spergel \& Haydn 2001]{jsh01}
Johnston, K. V., Spergel, D. N. \& Haydn, C. 2001, {\it in preparation}

\bibitem[Kauffmann, White \& Guiderdoni 1993]{cwg93}
	Kauffmann, G., White, S.D.M., Guiderdoni, 1993, MNRAS, 264, 201

\bibitem[Kauffmann \& White 1993]{kw93}
        Kauffman, G., White, S.D.M., 1993, MNRAS, 261, 921

\bibitem[Lacey \& Cole 1993]{lc93}
          Lacey, C., Cole, S., 1993, MNRAS, 262, 627

\bibitem[Lacey \& Cole 1994]{lc94}
      Lacey, C., Cole, S., 1994, MNRAS, 271, 676

\bibitem[Lequeux et al.\ 1996]{lequeux96}
	Lequeux, J., Fort, B., Dantel-Fort, M., Cuillandre, J.-C., \& 
	Mellier, Y. 1996, AA, 312, L1

\bibitem[Majewski et al. 1999]{m+99} 
Majewski, S.\ R., 
Siegel, M.\ H., Kunkel, W.\ E., Reid, I.\ N., Johnston, K.\ V., Thompson, 
I.\ B., Landolt, A.\ U.\ \& Palma, C.\ 1999, \aj, 118, 1709 
 
\bibitem[Majewski, Munn \& Hawley 1996]{mmh96} 
Majewski, S.\ R., Munn, J.\ A.\ \& Hawley, S.\ L.\ 1996, \apjl, 459, L73 
 
\bibitem[Malin \& Hadley 1997]{mh97}
Malin, D. \& Hadley, B. 1997, PASA, 14, 52

\bibitem[Mart{\'i}nez-Delgado et al, 2001]{delgado01} 
Mart{\'i}nez-Delgado, D., 
Aparicio, A., G{\'o}mez-Flechoso, M.\ A. \& Carrera, M. 2001, \apjl, 
{\it in press} 


\bibitem[Mateo 1998]{m98}
Mateo, M. 1998, \araa, 36, 435

\bibitem[Mateo, Olszewski \& Morrison 1998]{mom98}
	Mateo, M., Olzewski, E.W., Morrison, H.L. 1998, \apjl, 508, L55

\bibitem[Matthews, Gallagher \& van~Driel 1999]{matthews99}
	Matthews, L.D., Gallagher, J.S., van~Driel, W., 1999, AJ, 118, 2751

\bibitem[Morrison, Boroson \& Harding 1994]{mbh94}
Morrison, H.L., Boroson, T.A., \& Harding, P. 1994, \aj, 108, 1191

\bibitem[Neeser et. al.\ 2000]{neeser+00}
Neeser, M.J., Sackett, P.D., De~Marchi, G., \& Paresce, F. 2000, 
in preparation

\bibitem[Reitzel, Guhathakurta \& Gould 1998]{reitzel98} 
	Reitzel, D.B., Guhathakurta, P., Gould, A., 1998, AJ, 116, 707

\bibitem[Reshetnikov \& Sotnikova 2000]{rs00}
Reshetnikov, V.P. \& N.Y. Sotnikova, 2000, Astronomy~Letters, 26, 277

\bibitem[Rudy et al.\ 1997]{rudy97}
	Rudy, R. J., Woodward, C.  E., Hodge, T., Fairfield, S. W., 
	\& Harker, D. E. 1997, Nature, 387, 159

\bibitem[Sackett et al.\ 1994]{smhb94}
	Sackett, P.D., Morrison, H.L., Harding, P., Boroson, T.A., 1994, 
	Nature, 270, 441

\bibitem[Saha 1985]{saha85} 
	Saha, A., 1985, ApJ, 289, 310

\bibitem[Shang et al.\ 1998]{s+98}
Shang, Z. et al.\ 1998, \apjlett, 504, L23

\bibitem[Somerville \& Kolatt 1999]{sk99}
      Somerville, R.S., Kolatt, T.S., 1999, MNRAS, 305, 1

\bibitem[Somerville \& Primack 1999]{sp99}
         Somerville, R.S., Primack, J.R., 1999, MNRAS, 310, 1087

\bibitem[Somerville et al. 2000]{s+00}
     Somerville, R. S., Lemson, G., Kolatt, T. S., Dekel, A., 2000, 316, 479

\bibitem[Thoul \& Weinberg 1996]{tw96}
	Thoul, A.A., Weinberg, D.H., 1996, ApJ, 465, 608

\bibitem[Tremaine 1993]{t93}
Tremaine, S. 1993, in {\it Back to the Galaxy}, eds. S. S. Holt \& F. Verter
(New York: AIP), p. 599

\bibitem[Weil, Bland-Hawthorn \& Malin 1997]{weil97} 
Weil, M.\ L., Bland-Hawthorn, J.\ and Malin, D.\ F.\ 1997, \apj, 490, 664 

\bibitem[Weinberg 1993]{w93}
Weinberg, M. D. 1993, \apj, 410, 543

\bibitem[Whitelock, Irwin \& Catchpole(1996)]{whitelock96}
        Whitelock, P., Irwin, M., Catchpole, R. 1996, New Astron. 1, 57

\bibitem[Yanny et al. 2000]{sdss00a} 
Yanny, B.\ et al.\ 2000, \apj, 540, 825 

\bibitem[Zepf et al.\ 2000]{zepf+00}
Zepf, S.\ E., Liu, M.\ C., Marleau, F.\ R., Sackett, P.\ D.\ \&  
Graham, J.\ R.\ 2000, \aj, 119, 1701 

\bibitem[Zheng et al.\ 1999]{z+99}
Zheng, Z. et al. 1999, \aj, 117, 2757
 
\bibitem[Zinn 1985]{zinn85} 
	Zinn, R., 1985, ApJ, 293, 424

\end{thebibliography}
\end{document}